\documentclass[10pt,twocolumn]{article}

\usepackage{graphicx}
\usepackage{amsfonts}
\usepackage{amsmath}
\usepackage{amssymb}
\usepackage{mathrsfs} 
\usepackage{lipsum}
\usepackage{color}
\usepackage[squaren,Gray,cdot]{SIunits}
\usepackage{authblk}
\usepackage[margin=2cm]{geometry}
\usepackage{mathtools, cuted}
\usepackage{array,amsmath,booktabs}
\usepackage[font=normalsize, labelfont={bf,normalsize}]{subcaption}
\usepackage{bm}

\newcommand{\RH}{{\cal R}_H}
\newcommand{\LS}{L^{\star}}
\newcommand{\tS}{t^{\star}}
\newcommand{\slopeS}{\left. \frac{\partial h}{\partial t} \right|_{\star}}
\newcommand{\LSmoy}{\left\langle \LS \right\rangle}
\newcommand{\tSmoy}{\left\langle \tS \right\rangle}
\newcommand{\jd}{j_{\mathrm{d}}}
\newcommand{\je}{j_{\mathrm{e}}}

\title{\sf Influence of evaporation on soap film rupture}

\author[1,2]{Lor\`ene Champougny}
\author[1]{Jonas Miguet}
\author[1]{Robin Henaff}
\author[1]{Fr\'ed\'eric Restagno}
\author[1]{Fran\c{c}ois Boulogne}
\author[1]{Emmanuelle Rio}

\affil[1]{Laboratoire de Physique des Solides, CNRS, Univ. Paris-Sud, Universit\'e Paris-Saclay, Orsay 91405, France}
\affil[2]{MMN, Laboratoire Gulliver, CNRS, ESPCI Paris, PSL Research University, 10 rue Vauquelin, 75005 Paris, France.}

\begin{document}


\twocolumn[
    \begin{@twocolumnfalse}
        \maketitle
           \begin{abstract}
Although soap films are prone to evaporate due to their large surface to volume ratio, the effect of evaporation on macroscopic film features has often been disregarded in the literature.
In this work, we investigate experimentally the influence of environmental humidity on soap film stability.
An original experiment allows to measure both the maximum length of a film pulled at constant velocity and its thinning dynamics in a controlled atmosphere for various values of the relative humidity $\RH$.
At first order, the environmental humidity seems to have almost no impact on most of the film thinning dynamics. However, we find that the film length at rupture increases continuously with $\RH$.
To rationalize our observations, we propose that the film bursting occurs when the thinning due to evaporation becomes comparable to the thinning due to liquid drainage.
This rupture criterion turns out to be in reasonable agreement with an estimation of the evaporation rate in our experiment.
           \end{abstract}
	\vspace{0.5cm}
\end{@twocolumnfalse}]

\section{Introduction}

Bubble artists know very well that the soapy liquid they use to make giant bubbles needs to be adjusted depending on the weather conditions and in particular on the humidity of the atmosphere.
This empirical observation suggests that the evaporation of liquid from a soap film can have a direct impact on its stability.

More generally, the question of how and when a soap film ruptures is crucial in many different applied situations, ranging from water exchanges through aerosols production at the surface of oceans upon bubble bursting \cite{Monahan2001,Feng2014} to the control of foam stability in cosmetics or food industry \cite{Cantat2010}.
Foam coalescence is indeed a very drastic destabilization process for foams, which can be catastrophic for manufacturing light materials like foams concrete or very useful to recover the liquid phase after using a foam for nuclear decontamination.
However, studies on soap films, bubbles and foams' stability are most of the time performed at constant and measured humidity.
Experiments in which the humidity is systematically varied remain scarce in the literature  \cite{Li2010,Li2012}, certainly because understanding the rupture of soap films is already a challenge at fixed humidity \cite{Rio2014}.\\

The current picture of how soap films end up bursting can be decomposed into two main steps.
First, during the soap film’s lifetime, its thickness tends to decrease due to various mechanisms.
Gravity and capillary drainage generate liquid flows towards the bottom of the film \cite{Mysels1959} and the menisci \cite{Aradian2001,Howell2005}, respectively.
Marginal regeneration can also contribute to the film thinning through the rise of thin film patches generated near the menisci \cite{Stein1991,Aradian2001,Seiwert2017}.
Film thinning eventually results in the appearance of a 'black film', whose thickness is energetically stable due to the repulsion between the surfactant-laden interfaces of the film \cite{Cantat2010}.
However, instabilities due to surface concentration heterogeneities of surfactants can develop \cite{Bergeron1997} and lead to locally bare interfacial zones, which are very fragile and prone to burst due to thickness instabilities \cite{Vrij1964,Vrij1968,Lhuissier2009}.
The lifetime of a soap film thus depends on both the drainage dynamics and on instability mechanisms triggering the bursting \cite{Gennes2001,Rio2014}.

Evaporation can potentially impact both steps as it is an additional flux from the film to the atmosphere, which would tend to accelerate the thinning.
The nature of the stabilizing agents may influence the evaporation rate \cite{Schulman1962}, thus bringing more complexity to the boundary condition at the liquid/air interfaces \cite{Langevin2000a}.
If it is heterogeneous, the evaporation flux can also generate either temperature gradients and/or surface concentration gradients of chemical species.
Such gradients lead to Marangoni driven flows, which can on the one hand affect the drainage velocity and on the other hand either stabilize or destabilize the film \cite{Li2012, Pigeonneau2012}. \\

The goal of this article is to explore the impact of evaporation on soap film rupture.
We propose to address this question in an original way by measuring the thinning dynamics and maximum length of soap films, which rupture during their generation in a humidity-controlled atmosphere.
The corresponding experimental setup is described in section \ref{sec:setup}.
Our experimental results are then presented in section \ref{sec:results}, where we show that, surprisingly, the film thinning dynamics is not affected by the humidity up to the bursting.
Yet, the film maximum length is found to be a function of the environmental humidity.
In order to rationalize this last finding, we propose that the film rupture occurs when the drainage rate becomes close to the evaporation rate, as discussed in section \ref{sec:discussion}.

\section{Experimental Setup}\label{sec:setup}
\begin{figure}[ht]
	\centering
    \includegraphics[width=8cm]{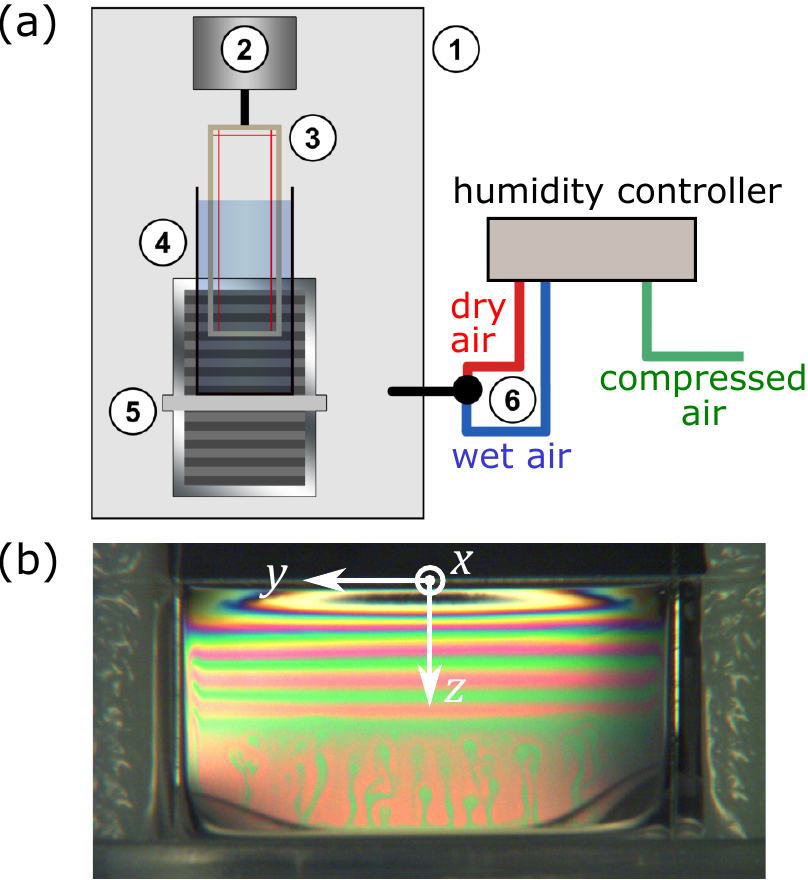}
	\caption{(a) Experimental setup used to generate and study soap films in a controlled atmosphere. \textcircled{1} Closed box, \textcircled{2} force sensor, \textcircled{3} static frame supported by the force sensor, \textcircled{4} plastic tube containing the solution supported by \textcircled{5} the translation table, \textcircled{6} humidity control device. (b) Photograph of a soap film during its generation at $V=20~\milli\meter\per\second$, in an environmental humidity $\RH = 80~\%$.
	}
    \label{fig:Setup}
\end{figure}
%
\subsection{Film generation}
The experimental protocol used to generate soap films consists in withdrawing at a constant velocity $V$ a vertical frame out of a reservoir containing a soapy solution.
The corresponding experimental setup is pictured in Fig.~\ref{fig:Setup}(a)).

The main frame, made with a 3D-printer, is a rectangle of dimensions $20 \times 90 ~\milli \meter^2$.
On this main frame, two vertical and one horizontal nylon threads (Nanofil, Berkley) of diameter $140~\micro\meter$ are glued.
This secondary frame is the support of the free standing soap film.

A cylindrical reservoir of $2.8~\centi\meter$ in diameter contains a solution of TTAB (tetradecyl trimethylammonium bromide, purchased from Sigma-Aldrich and used as received) at a fixed concentration of $5~\gram\per\liter$, corresponding to approximately 4 times the critical micellar concentration ($\mathrm{cmc} = 3.6~\milli\mole\per\liter$ \cite{Bergeron1997}).
This reservoir is displaced vertically at a controlled velocity $V$ using a motorized linear stage (Newport UTS150CC) coupled to a motion controller (Newport SMC100CC).
The displacement of the stage is recorded in time and used to determine the height of the free standing soap film $L(t)$ in time.
%
\subsection{Film characterization}
A force sensor located at the top of the main frame allows to detect automatically the film rupture, as developed in reference \cite{Saulnier2014}.
The film lifetime $\tS$ is thus determined and its maximum length $L(\tS) = \LS$ is deduced from the displacement measurements.
In the following, we choose to express our results in terms of $\LS$ since $\tS$ can be rather misleading when varying the velocity.
Indeed, $\tS$ becomes quite large at small velocity just because the pulling dynamics is slow, whereas the film is very unstable.
The determination of $\LS$ requires an accurate determination of the position of the surface of the liquid reservoir.
The latter is measured before each set of experiment (for a given pulling velocity and at a given humidity) because it can vary with time due to evaporation.
The detection is done by approaching slowly ($V=0.5~\milli\meter\per\second$) the top wire to the interface until a contact is observed.
The vertical position is then reported and considered as the zero position for the corresponding experiment.
The error on this reference position is typically $<0.5~\milli\meter$.\\

The soap film thickness is measured locally using a reflectometry technique.
An optical fiber (IDIL, France) with a lens allows to focus a white light spot on the soap film.
The reflected light spectrum is collected by a second optical fiber (IDIL, France) and measured by a spectrometer (USB 400, Ocean Optics) in the wavelength range $400 - 1000~\nano\meter$.
The reflected intensity $I_{\mathrm{r}}$ normalized by the incident intensity $I_0$ can be formally expressed for each wavelength $\lambda$ as
\begin{equation}
\frac{I_{\mathrm{r}}(\lambda)}{I_0(\lambda)}=\frac{\sin^2(\frac{2\pi n h}{\lambda})}{\left(\frac{2n}{n^2-1}\right)^2+\sin^2(\frac{2\pi n h}{\lambda})},
\label{eq:interferences}
\end{equation}
where $h$ is the film thickness and $n$ the optical index of the solution.
The film thickness $h$ is obtained by fitting the experimental spectrum with Eq. \ref{eq:interferences}.
This procedure yields accurate values for the film thickness as long as the spectrum features at least one oscillation, which corresponds to films thicker than about $200~\nano\meter$.

For thinner films, we use the method derived by Scheludko \cite{Sheludko1967} in 1967.
When the order of interference is zero (\textit{i.e.} for $h < \lambda/2n$), the relationship between $I_{\mathrm{r}}$ and $h$ becomes bijective for each wavelength $\lambda$.
For a given wavelength, Eq. \ref{eq:interferences} can thus be inverted in order to obtain directly the thickness $h$ as a function of the reflected intensity.
In practice, we extract the thickness for $7$ different wavelengths in the range $550-850~\nano\meter$, check that they yield approximately the same value for $h$ and average the thickness over these different wavelengths.
%
\subsection{Environmental regulation}
The film generation setup described above is enclosed in a box of dimensions $40 \times 50 \times 50 ~\centi\meter^3$ in which the humidity is regulated using a home-made controller (pictured in Fig.~\ref{fig:Setup}(a)).
A PID controller based on an Arduino Uno and a humidity sensor (Honeywell HIH-4021-003) positioned far from the evaporating surface allow to inject the adequate proportions of dry and moist air to reach the target humidity in the box.
Dry air is produced by circulating ambient air with an air pump (Tetra APS 300) in a container filled with desiccant made of anhydrous calcium sulfate (Drierite).
Moist air is obtained by bubbling air in water.
To achieve measurements at $\RH \approx 100$ \%, we saturate the atmosphere before starting the regulation by paving the box with damp sponges.
The whole setup yields a typical uncertainty of $\pm 1~\%$ on the relative humidity.
The temperature within the box is kept constant at $T=20 \pm 1~\degree\Celsius$.
%
\section{Experimental results} \label{sec:results}
%
\subsection{Rupture length vs velocity and relative humidity}
\begin{figure}[!ht]
\centering
	\includegraphics[width=8cm]{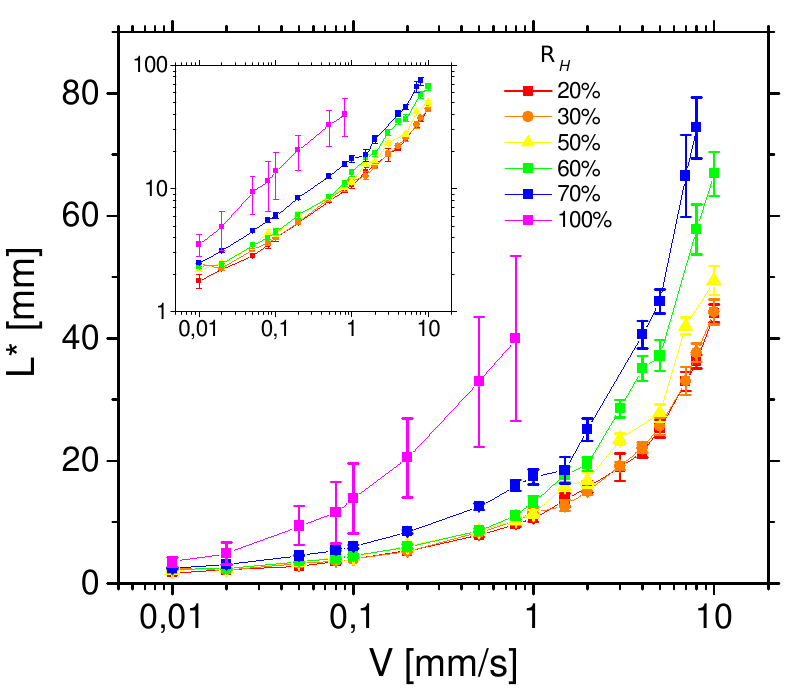}
    \caption{Maximum length $\LS$ of soap films as a function of the pulling velocity $V$ for different relative humidities $\RH$.
    Each point corresponds to an average over at least 20 measurements and the error bar represents the standard deviation.
    The inset shows the same data in a log-log scale.
    }
    \label{fig:LStarVsV}
\end{figure}
\begin{figure}[!ht]
    \includegraphics[width=8cm]{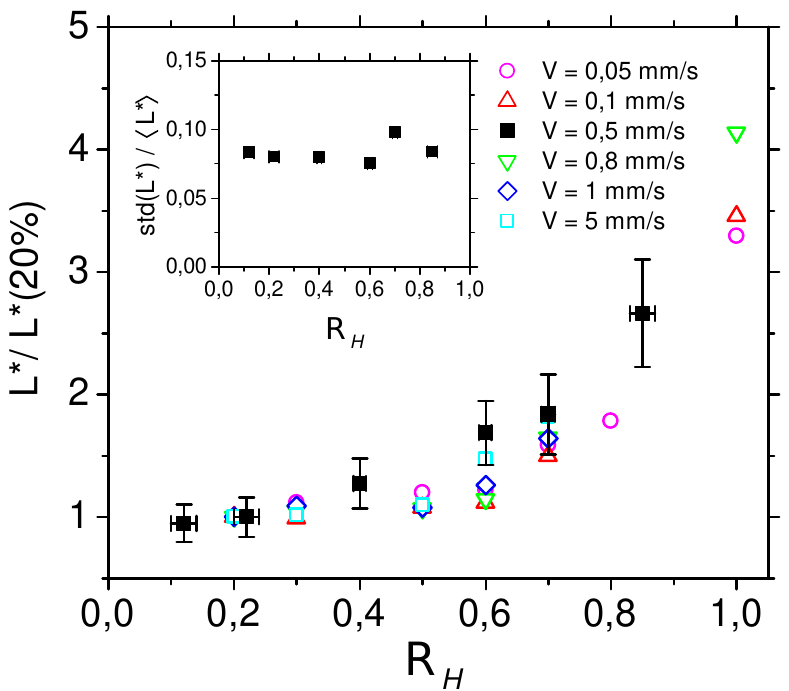}
    \caption{For different pulling speeds $V$, we plot the maximum length $\LS$ of soap films, normalized by its value at $\RH = 20~\%$, as a function of the environmental humidity $\RH$.
    Empty symbols correspond to averages over about 20 measurements, extracted from Fig. \ref{fig:LStarVsV}.
    Solid symbols correspond to averages over at least 200 measurements and the error bars represent the standard deviation of the corresponding data.
    For this last set of data ($V=0.5~\milli\meter\per\second$), the inset shows the evolution of the ratio between the standard deviation $\mathrm{std}(\LS)$ and the average value $\LSmoy$ with the environmental humidity.
    }
    \label{fig:LStarVsRH}
\end{figure}
In Fig.~\ref{fig:LStarVsV}, we report the rupture lengths $\LS$ as a function of the pulling velocity $V$ for various relative humidities $\RH$.
We observe that $\LS$ increases with the pulling velocity roughly as a power law, with exponents in the range $0.45-0.55$ for all the values of relative humidity we probed (see inset of Fig.~\ref{fig:LStarVsV}).
The trend of the data displayed in Fig.~\ref{fig:LStarVsV} is consistent with the work of Saulnier \textit{et al.} \cite{Saulnier2014}, where the maximum length of soap films $\LS$ was measured at ambient humidity ($\RH \sim 35~\%$) for different surfactants and concentrations.

In Fig.~\ref{fig:LStarVsRH}, the empty symbols show the same data as in Fig.~\ref{fig:LStarVsV} but plotted as a function of the relative humidity $\RH$ for various pulling speeds $V$.
In order to compare the data for different velocities, the maximum length $\LS$ is normalized by its value at $\RH = 20~\%$ for each $V$.
We also performed an additional set of experiments at $V=0.5~\milli\meter\per\second$ and measured the maximum length of at least 200 films for each value of $\RH$.
The corresponding averaged data are represented in Fig.~\ref{fig:LStarVsRH} by solid symbols and the error bars show the standard deviation of the measurements.
In addition, the standard deviation $\mathrm{std}(\LS)$ divided by the mean value $\LSmoy$ of the distribution of film heights at $V=0.5~\milli\meter\per\second$ is shown in the inset of Fig.~\ref{fig:LStarVsRH}.

For a given pulling velocity, the film maximum length $\LS$ increases with the relative humidity $\RH$, as could already be observed in Fig.~\ref{fig:LStarVsV}.
This increasing behavior turns out to be nonlinear.
When $\RH$ tends to $0$, the rupture length $\LS$ seems to tend towards a constant value, which depends on the velocity.
On the contrary, $\LS$ rises sharply when approaching $\RH=1$.
This behavior appears to be quite independent of the pulling speed, since the data obtained with $V$ varying over two orders of magnitude collapse onto a single mastercurve when $\LS$ is normalized by its low-humidity value (taken at $\RH = 20~\%$).

Interestingly enough, the ratio $\mathrm{std}(\LS) / \LSmoy$ measured for $V=0.5~\milli\meter\per\second$ does not depend on the environmental humidity in the range $\RH = 10-85~\%$, as shown in the inset of Fig.~\ref{fig:LStarVsRH}.
This suggests that the physical mechanism at the origin of the stochastic nucleation of a hole in the black film is not affected by the environmental humidity.
%
\subsection{Experimental thinning dynamics}
\begin{figure}[ht!]
    \includegraphics[width=8cm]{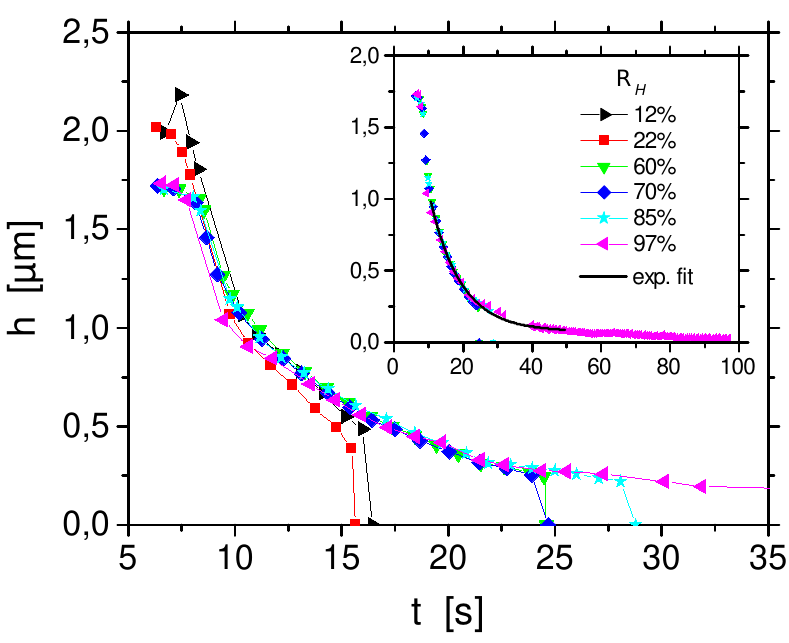}
    \caption{Time evolution of the film thickness $h$ for various relative humidities $\RH$ and a fixed pulling speed $V=0.5~\milli\meter\per\second$.
    The thickness is measured at the top of the film, about $1~\milli\meter$ below the horizontal thread.
    The inset shows the same data, along with the long-time evolution of the film thickness at $\RH = 97~\%$ and an exponential fit of the data in the range $10 - 50~\second$.
    }
    \label{fig:data_drainage}
\end{figure}
A possible explanation for the enhanced stability of soap films at high relative humidities is that the reduction of evaporative effects significantly slows down their thinning dynamics.
To test this hypothesis, we measured the variation of the thickness $h$ with time $t$ at the top of the film for various relative humidities $\RH$.
These experiments were performed at a fixed velocity of $V=0.5~\milli\meter\per\second$, as for the maximum length measurements shown in Fig.~\ref{fig:LStarVsRH} (solid symbols).

The results are plotted in Fig.~\ref{fig:data_drainage}, where the reference time $t=0$ is the birth time of the film ($\pm~1~\second$), defined as the time when the horizontal thread crosses the surface of the bath.
These measurements show a continuous decrease of the film thickness, which follows the same trend with no perceptible effect of the relative humidity, until the film breaks.
On the contrary, the rupture is marked by a sharp decrease of the thickness to zero at a time which depends strongly on $\RH$.
Note that the rupture dynamics could not be resolved in time with our thickness measurement method.
To put it in a nutshell, the film thinning seems to be insensitive to the environmental humidity, while the bursting is.
%
\section{Data analysis and discussion} \label{sec:discussion}
%
\subsection{Criterion for film bursting}
To rationalize the thinning curves shown in Fig.~\ref{fig:data_drainage}, we propose to write a simplified 1D model to describe the variation of the film thickness in time.
Denoting $x$ the direction normal to the film, $y$ the coordinate along the film width and $z$ the vertical coordinate (see Fig.~\ref{fig:Setup}), we assume that the thickness $h(z, t)$ is invariant along $y$.
This hypothesis is supported by the observation of horizontal fringes in Fig.~\ref{fig:Setup}(b).
Therefore, the time evolution of the film thickness is attributed to a combination of a vertical liquid drainage and a mass loss due to evaporation across the liquid-vapor interfaces.
Thus, we write
\begin{equation}\label{eq:thinning}
	\frac{\partial h}{\partial t}(z, t, \RH) = -\jd(z,t) - 2 \je(z,\RH),
\end{equation}
where $\jd(z,t)$ corresponds to the liquid drainage in the film along the vertical $z$-axis, which depends on time, and $\je(z,\RH)$ is the local evaporative flux, that we assume to be constant.
The factor $2$ accounts for the two liquid-vapor interfaces.
The similar trend of the film thinning for different humidities suggests that the liquid drainage is decoupled from the evaporation, such that the term $\jd(z,t)$ is rendered by the measurements at $\RH=100~\%$, where the evaporative flux is zero.
If some Marangoni gradients were generated by temperature or surfactant concentration gradients as proposed earlier \cite{Li2012,Pigeonneau2012}, they would modify the boundary condition at the interfaces of the film and therefore affect the drainage differently depending on $\RH$.
This effect thus appears to be negligible in our experiment.

Based on Eq.~\eqref{eq:thinning}, we propose that two regimes can be distinguished.
For $t \ll t^\star$, the influence of evaporation is negligible and the film thinning is dominated by drainage.
Thus, $\frac{\partial h}{\partial t} \sim -\jd(z,t)$ and the thinning data are on a master curve, independent on $\RH$.
Since $\jd(z,t)$ is a decreasing function of time, the evaporation and the drainage fluxes will eventually become of the same order of magnitude.
We hypothesize that this is responsible for the film bursting, occurring for $t \sim t^\star$,
The corresponding rupture criterion can be expressed as a scaling
\begin{equation}\label{eq:scaling_flux}
	\jd(0,t^\star) \sim 2 \je( 0,\RH),
\end{equation}
where the fluxes are evaluated at the top of the film ($z\approx 0$), where the bursting takes place \cite{Saulnier2014}.
%
\subsection{Estimation of the evaporation rate $\je$}
In order to test the validity of the rupture criterion Eq.~\ref{eq:scaling_flux}, let us first estimate the evaporation rate $\je$.
For a vapor concentration field $c(x,y,z)$, the evaporative flux is defined as
\begin{equation}\label{eq:def_je}
	\je = - \frac{ {\cal D} }{\rho} \left.\frac{\partial c}{\partial x}\right|_{x=0},
\end{equation}
where ${\cal D}=2\times 10^{-5}~\meter^2\per\second$ is the diffusion coefficient of water and $\rho$ its density.
To calculate this evaporative flux, the concentration field must be determined first.
Indeed, the transport of water in the vapor phase can be either diffusive \cite{Langmuir1918} or convective \cite{Shadizadeh-Bonn2006,Dehaeck2014} depending on the competition between the buoyancy of the vapor and the viscosity of the gas.
For water, the characteristic lengthscale of the evaporating interface above which convection becomes significant is typically $5-10~\milli\meter$ \cite{Dollet2017}.

In our experiments, both the film and the reservoir evaporate and their sizes are of the order of the centimeter.
Thus, we can expect that convection has a non-negligible effect.
However, taking into account the convective flow is particularly difficult as it depends strongly on the geometry \cite{Dollet2017}.
Consequently, we evaluate the evaporative flux from a scaling analysis of Eq. \ref{eq:def_je} by introducing a characteristic lengthscale ${\cal L}$ of the vapor concentration gradient, \textit{i.e.}
\begin{equation}\label{eq:scaling_je}
	\je \sim \frac{ {\cal D} }{\rho}  \frac{c_s - c_\infty}{ {\cal L}},
\end{equation}
where $c_s$ and $c_\infty$ are respectively the mass concentration of the saturated vapor and far from the evaporating liquid.
The mass concentration far from the soap film is $c_\infty = c_s \RH $.

At room temperature, the saturated pressure of water is $P_s \approx 2.3~\kilo\pascal$ \cite{Tennent1971}.
Therefore, the saturated mass concentration is $c_s = P_s {\cal M} / (R T)$, with the water molecular weight ${\cal M} = 18~\gram\per\mole$, the ideal gas constant $R$ and $T$ the temperature.
In appendix \ref{apdx:evap}, we checked that the presence of TTAB molecules at a concentration of 4 times the cmc does not modify significantly the solution activity.
Thus, we will henceforth consider that the evaporation kinetics of our soap solution is that of pure water.

As no precise model is derived, we do not claim to be fully predictive on the evaporation kinetics.
As stated before, the precise modeling of the vapor concentration field surrounding the reservoir and the withdrawn film is particularly challenging.
For film lengths much larger than the radius of the reservoir, we would expect that the convective evaporation satisfies the dynamics for vertical films \cite{Boulogne2018b}.
In the opposite limit of small films, the evaporation rate would be mainly set by the vapor surrounding the circular reservoir \cite{Dollet2017}.
Here, we are in an intermediate situation, where the characteristic lengthscale of the withdrawn films at rupture $\LSmoy$ is typically between 8 to 20 mm (Fig.~\ref{fig:LStarVsRH}), comparable to or slightly larger than the reservoir radius.
Thus, a direct comparison with more advanced modeling available in the literature\cite{Boulogne2018b} is not possible due to the different boundary condition set by the soap solution reservoir, which also evaporates.
%
\subsection{Comparison to experimental data} \label{sec:comp}

The rupture criterion given in Eq.~\ref{eq:scaling_flux} involves the drainage flux $\jd$ evaluated at the rupture time $\tS$.
However, the theoretical modelling of the drainage of a vertical film pulled at constant velocity is a challenging task in itself \cite{Heller_phd,Champougny2017}, which lies beyond the scope of this paper.
Since we do not have access to the drainage flux $\jd$ directly, we will adopt a more phenomenological approach, where we measure the instantaneous slope of the thinning curve just before rupture, defined as $\slopeS (\RH) = \frac{\partial h}{\partial t} (0, \tS, \RH)$.

Combining Eqs.~\ref{eq:thinning} and \ref{eq:scaling_flux}, we can express the rupture criterion in terms of the thinning rate at bursting $\slopeS$ as
\begin{equation}\label{eq:scaling_flux_reformule}
	- \slopeS (\RH) \sim 4 \je( 0,\RH),
\end{equation}
where $\je$ varies linearly with $1-\RH$ according to Eq.~\ref{eq:scaling_je}.

\begin{figure}
    \includegraphics[width=8cm]{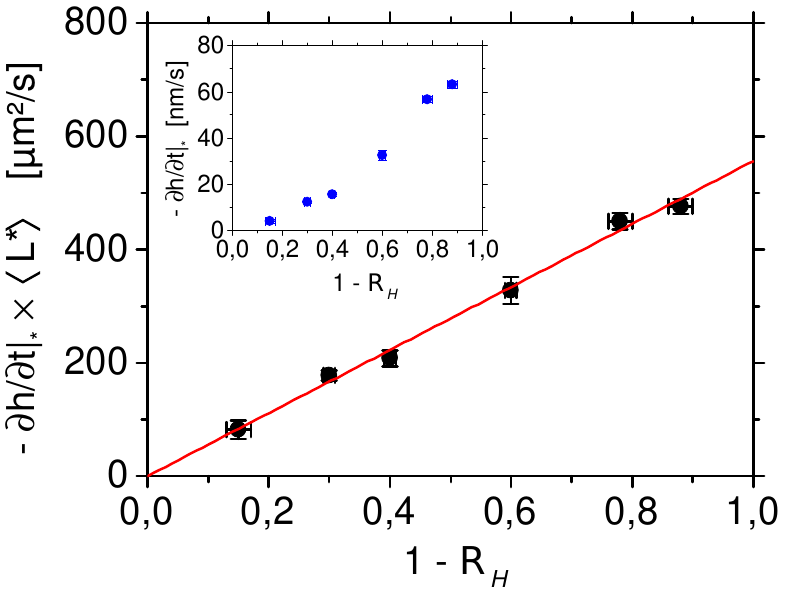}
    \caption{
    The main plot shows the variation of $-\slopeS \times \LSmoy$ as a function of $1-\RH$.
    The red solid line is a linear fit forced to pass through the origin.
    The corresponding slope is found to be $556 \pm 10~\micro\meter^2\per\second$.
    The inset shows the raw data $\slopeS$ as a function of $1-\RH$.
    }
    \label{fig:thinning_vs_RH}
\end{figure}

In order to extract $\slopeS$ from the experimental data, we cannot simply derivate the experimental curve, which would add too much noise.
Instead, we fit the master curve $h(t)$ observed for the film thinning (Fig.~\ref{fig:data_drainage}) with \textit{ad-hoc} analytical functions of time.
The derivative $\frac{\partial h}{\partial t}$ can then be calculated analytically and only depends on time since $\RH$ has no impact on this master curve.
For each humidity, $\slopeS$ is calculated by evaluating this analytical function at $\tSmoy(\RH)=\LSmoy(\RH)/V$, where $\LSmoy$ is obtained from Fig.~\ref{fig:LStarVsRH}.
The extracted values of $\slopeS$ are plotted in the inset of Fig.~\ref{fig:thinning_vs_RH} as a function of $1-\RH$.
We checked that the function used to fit $h(t)$ can be chosen arbitrarily, as long as it describes well the data.
More details on the various fitting functions tested can be found in appendix \ref{apdx:fit}.
For the sake of illustration, a decreasing exponential fit of the data is presented in the inset of Fig.~\ref{fig:data_drainage}.

Our experimental results on the thinning rate (Fig.~\ref{fig:thinning_vs_RH}) indicate a better linear agreement against $1-\RH$ for a characteristic lengthscale ${\cal L}$ varying as $\LSmoy$.
Substituting this presumption in Eq.~\eqref{eq:scaling_flux_reformule}, we have
\begin{equation}\label{eq:scaling_je_v2}
	-\slopeS \times \LSmoy \, \sim \, 4 \, \frac{ {\cal D} c_s}{\rho} (1 - \RH).
\end{equation}
This equation is tested in Fig.~\ref{fig:thinning_vs_RH} where the data are fitted with a linear relationship, which is particularly convincing.
The linear fit yields a prefactor $4 {\cal D} c_s/\rho \simeq 5.6 \times 10^2~\micro\meter^2\per\second$, which is comparable to the value estimated for water $4 {\cal D} c_s/\rho \simeq 12 \times 10^2~\micro\meter^2\per\second$.

The phenomenological rupture criterion \eqref{eq:scaling_je_v2} allows us to recover the correct trend for the slope $\slopeS$ just prior to rupture as a function of $\RH$.
This simple modeling sheds light on how the environmental humidity $\RH$ can affect the maximum length of soap films $\LS$ without significantly altering the overall thinning dynamics of the film in its early life.
However, this approach raises several questions.
Indeed, the sharpness of the transition from a slow, $\RH$-independent drainage regime to a fast, $\RH$-dependent rupture regime seems quite surprising.
Moreover, the rupture criterion involves the typical length scale $\mathcal{L}$ of the vapor concentration gradient, which is \textit{a priori} unknown.
It is still an open question to know why this length scale is reasonably approximated by the film maximum length $\LS$.
Future experiments with different boundary conditions, that would be easier to describe theoretically, will hopefully help making progress on these questions.
%
\section{Conclusion}

An automatized experiment was developed to measure the maximum length of soap films generated in a humidity-controlled atmosphere.
The film length was found to increase nonlinearly with the relative humidity $\RH$.
By carefully measuring the thinning dynamics at the top of the film, we showed that in our experiments the thinning dynamics is almost not affected by the evaporation.

This important observation led us to make the hypothesis that evaporation becomes significant only very close to rupture.
We thus proposed a phenomenological rupture criterion which is that the film breaks when the mass loss due to evaporation becomes of the order of the mass loss due to drainage at the top of the film.
Following this hypothesis, we extracted from our drainage data the value of the mass loss at the top of the film just before rupture, which appears to be in reasonable agreement with the value expected for diffusion-driven evaporation, provided the typical length scale of the concentration gradient is set by the maximum length of the film.

Interestingly, we did not observe any experimental signature of thermal or solutal Marangoni stresses that may be induced by inhomogeneous evaporation in the configuration of our experiment.
The identification of configurations in which these possible Marangoni flows become non negligible in soap films would certainly deserve interest.
This work opens the route to more investigations concerning the impact of evaporation on foam films stability and, more generally, on foam stability.
We also expect that these results will be valuable for future theoretical developments on soap film instability leading to rupture.
%
\appendix
\section{Evaporation kinetics of TTAB solution vs water} \label{apdx:evap}
\begin{figure}[!ht]
     \includegraphics[width=8cm]{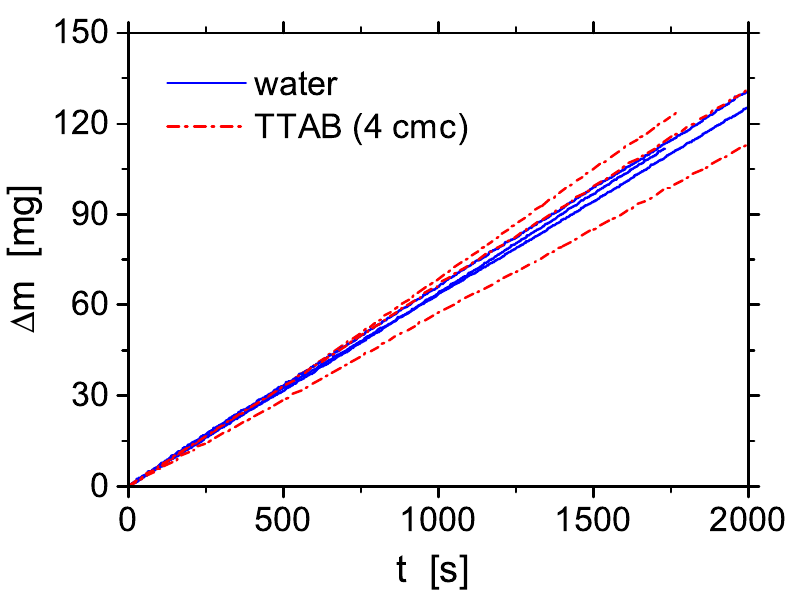}
     \caption{Cumulated evaporated mass $\Delta m$ as a function of time for pure water and a TTAB solution of concentration 4 cmc, measured in an environmental humidity $\RH=50~\%$.}
     \label{fig:evap_kinetics}
\end{figure}
In order to check that the presence of surfactant does not modify significantly the solution activity, we compared the evaporation rates of water and of a solution of TTAB at a concentration of 4 cmc, contained in Petri dishes ($5.7~\centi\meter$ in diameter) filled up to the rim.
In each case, the cumulative mass loss $\Delta m$ is measured as a function of time $t$ in an environmental humidity $\RH=50~\%$.
Three measurements were carried on both for water and the solution of interest, and the results are shown in Fig.~\ref{fig:evap_kinetics}.

The average mass fluxes, estimated from the slopes of the curves, are $64 \pm 1~\micro\gram\per\second$ and $64 \pm 5~\micro\gram\per\second$ for water and TTAB solutions respectively.
The larger dispersion in the case of TTAB deserves further investigation and may be due to the shape of the meniscus at the rim which is not controlled finely and could result in inhomogeneities of evaporation.
However, the average mass fluxes are very close, which allows us to safely consider that the activity of the TTAB solution can be assimilated to that of water.
%
\section{Robustness of the estimation of the evaporation rate} \label{apdx:fit}
\begin{figure}[!ht]
     \includegraphics[width=8cm]{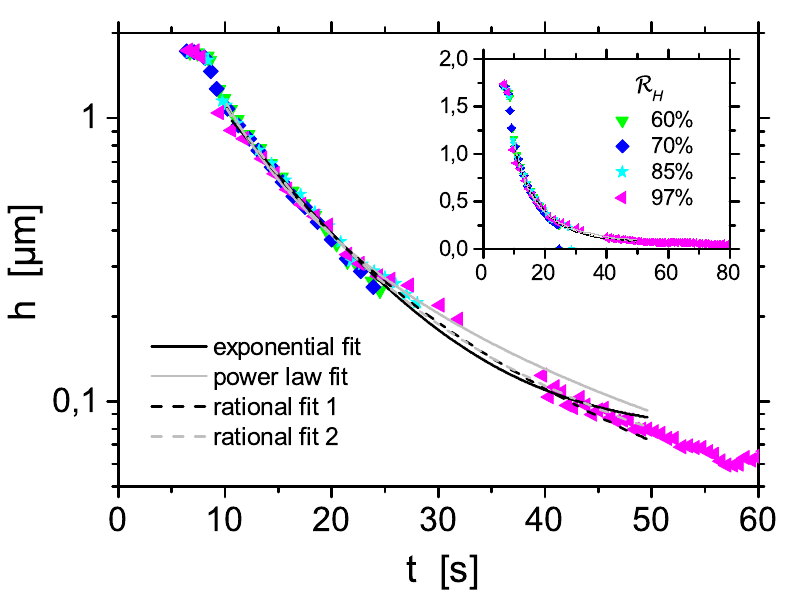}
     \caption{The drainage data of Fig.~\ref{fig:data_drainage} are replotted in log (main graph) and linear scales (inset), along with fits performed with four different functional forms, as developed in Table~\ref{tab:fit_func}.}
     \label{fig:fit_all_functions}
\end{figure}
\begin{table}[!ht]
\centering
\begin{tabular}{l
        >{$\displaystyle}l<{$}     
        l}
Name & \text{Functional form}~h(t) & Parameters   \\
\hline
exponential & A_1 \exp \left( -\frac{t}{t_1}\right) + h_0 & $A_1$, $t_1$, $h_0$   \\
power law & at^b \quad (b<0) & $a$, $b$   \\
rational 1 & \frac{1}{a + bt^c} & $a$, $b$, $c$   \\
rational 2 & A_2 + \frac{A_1 - A_2}{1 + (t/t_0)^p} & $A_1$, $A_2$, $t_0$, $p$  \\
\end{tabular}
\caption{Functional forms and corresponding adjustable parameters used to fit the drainage data.}
\label{tab:fit_func}
\end{table}
In paragraph \ref{sec:comp}, the drainage dynamics of the films are empirically accounted for by fitting \textit{ad-hoc} functions to the experimental data.
In order to ensure the robustness of this approach, we tested four different functional forms $h(t)$, as developed in Table \ref{tab:fit_func}.
The corresponding fits to the experimental data are presented in log and linear scales in Fig.~\ref{fig:fit_all_functions}.

The slope just before rupture $\slopeS (\RH) = \frac{\partial h}{\partial t} (0, \tS, \RH)$ is then obtained analytically for each of the functional forms.
The data points presented in Fig.~\ref{fig:thinning_vs_RH} are averages of the values obtained for the four different functions presented in Fig.~\ref{fig:fit_all_functions} and the vertical error bars represent the standard deviation.
%
\section*{Acknowledgments}

We are very grateful to J\'er\'emie Sanchez and Vincent Klein for the research and development of the humidity controller.
We also thank M\'elanie Decraene for her assistance.
%
%
\bibliography{biblio}
\bibliographystyle{unsrt}
%
\end{document}